\documentclass{article}
\usepackage{specom2006,amssymb,amsmath,epsfig}
 \usepackage{html,url,graphicx}
 \latex{\renewcommand{\htmladdnormallink}[2]{#1 (\url{#2})}}
\ninept
\def\reg{{\rm\ooalign{\hfil
     \raise.07ex\hbox{\scriptsize R}\hfil\crcr\mathhexbox20D}}}

\newcommand{\figref}[1]{Figure~\ref{#1}}

\title{Synonym Search in Wikipedia: Synarcher}

\name{Andrew A. Krizhanovsky}

\address{Computer-Aided Integrated Systems Laboratory \\
St.Petersburg Institute for Informatics and Automation of the Russian Academy of Sciences, \\
St.-Petersburg, Russia \\
{\small \tt aka@mail.iias.spb.su}
}
\begin{document}
\maketitle
\begin{abstract}
The program Synarcher for synonym (and related terms) search in the text corpus of special structure (Wikipedia) was developed. 
The results of the search are presented in the form of graph. 
It is possible to explore the graph and search for graph elements interactively. 
Adapted HITS algorithm for synonym search, program architecture, and program work evaluation with test examples are presented in the paper. 
The proposed algorithm can be applied to a query expansion by synonyms (in a search engine) and a synonym dictionary forming.\end{abstract}

\section{Introduction}

Growing number of documents, 
appearance of new auxiliary structures and metadata linked to documents 
enable one to adapt known algorithms and to propose new ones for more precise search. 
The similarity search problem includes search of similar text documents, semantically related words, and graph similar vertices. 
There are enough algorithms for solving this problem, 
e.g., the ``hypertext induced topic selection''(HITS) algorithm proposed by Kleinberg \cite{kleinberg99}, 
PageRank algorithm \cite{brin_pagerank98}, Vector Space Model, Latent Semantic Indexing and others.

The paper presents an adapted HITS algorithm and its implementation as a program 
for synonyms and related terms search in a corpus with hyperlinks and categories (Wikipedia).
Since the developed algorithm uses a link structure (hyperlinks in texts), it is language independent.

In order to select text resources and software libraries the following criteria were taken into account. 
\begin{itemize}
\item It should be open, or at least free.
\item The developed program and used libraries should be operating system independent.
\item Some library for visualization of synonym search result is needed.
\end{itemize}

The Wikipedia (as the text corpus) and 
\htmladdnormallinkfoot {Touch\-Graph Wiki\-Browser} {http://www.touchgraph.com} 
(visualization library) were selected. 
The proposed synonym search algorithm and the developed program 
\htmladdnormallinkfoot {Synarcher} {http://sourceforge.net/projects/synarcher} 
could be used in order to extend query in a search engine, or as an assistant for forming a dictionary of synonyms.



\subsection{Wiki resource: Wikipedia}

Wiki is a type of sites that provide a simple way to add and modify pages for users. 
It is specially designed for \htmladdnormallinkfoot{collaborative work} 
    {http://en.wikipedia.org/wiki/Wiki}. 
Wiki is also a software package at server side enabling one to add/modify Internet pages content via Internet browsers. 
Wiki language supports hyperlinks (to create links between wiki pages), 
it is more human readable than HTML and more safe (there are no JavaScript and Cascading Style Sheets).

The wiki resource Wikipedia is used in this work as a text corpus. 
\htmladdnormallinkfoot {Wikipedia} {http://wikipedia.org} 
is a free online encyclopaedia in English, Russian, etc.

The research and discussion of Wikipedia resource are presented in \cite{holloway05}. 
\htmladdnormallinkfoot{A Wikipedia article} 
    {http://en.wikipedia.org/wiki/Wikipedia:What\_is\_an\_article} 
is defined as a page 
that has encyclopedic or almanac-like information  
("almanac-like" being lists, timelines, tables or charts). 
Wikipedia text corpus has the following features valuable for synonym search:
\begin{itemize}
\item The texts classification is defined via categories \cite{voss06}. 
      Article authors select and assign most suitable categories to the article. 
      It is possible to create a new category and link it to other relevant categories.
\item The encyclopedia contains a lot of articles related to different topics, 
      related to modern topics, since the encyclopedia is updated every day.
\end{itemize}

\subsection{Corpus requirements}
\label{corpus}

The set of documents constitutes a corpus of the documents. 
Hyperlinks and categories are used to search for similar texts in the corpus. 
The texts in the corpus are characterized as follows:
\begin{itemize}
\item [1.] Text documents include a set of keywords, if keywords are not given, the document title is considered as a set of keywords.
\item [2.] The documents refer to each other via hyperlinks. Every document has a set of out-links (hyperlinks to documents this document refers to) and in-links (hyperlinks this document is referred by). 
The hyperlinks are assigned by experts.
\item [3.] Every document belongs to a set of categories/topics. Belonging of a document to a category is defined by experts. 
Categories make up a tree-like structure, 
thus every category has a parent category (except the root) and child categories (except the leaves).
\end{itemize}

Wikipedia satisfies corpus requirements. 
The search of synonyms within such corpus could be presented in a strict mathematical formulation.

\section{HITS algorithm}

HITS algorithm searches Web pages that are relevant to a given query. 
One of alternatives to the HITS algorithm is the Page rank algorithm \cite{brin_pagerank98} incorporated in Google.com. 
\htmladdnormallinkfoot {Google description of PageRank is suitable for HITS} {http://www.google.com/technology}: 
\begin{quotation} 
PageRank relies on the uniquely democratic nature of the web by using its vast link structure as an indicator of an individual page's value. In essence, Google interprets a link from page A to page B as a vote, by page A, for page B. But, Google looks at more than the sheer volume of votes, or links a page receives; it also analyzes the page that casts the vote. Votes cast by pages that are themselves ``important'' weigh more heavily and help to make other pages ``important''.
\end{quotation}

The notion of authority is used in HITS. Authority has meaning only in the context of a particular query topic. Link structure is used in HITS to identify page authority, since hyperlinks encode a considerable amount of latent human judgment.
"\emph{Authoritative pages} relevant to the initial query should not only have large in-degree (hyperlinks this document is referred by); 
since they are all authorities on a common topic, there should also be considerable overlap in the sets of pages that point
to them. Thus, in addition to highly authoritative pages, it is needed to find what could be called \emph{hub pages}: 
these are pages that have links to multiple relevant authoritative pages" \cite{kleinberg99}.

It is proposed to use link structure in HITS in order to satisfy similar-page queries.
Using the notion of hubs and authorities, it is possible to solve the issue of page similarity.
Given a page \(s\) that is of interest, 
the authorities in the local region of the link structure near page \(s\) can potentially serve 
as a broad-topic summary of the pages related to \(s\). 

The idea of link structure usage is adapted for synonym search in Wikipedia.

\section{Adapted HITS algorithm}

HITS algorithm was adapted in order to use the additional characteristics of documents (pages) provided by the corpus.

Since pages include a set of keywords\footnote{See 
      the 1\(^{st}\) characteristic of corpus documents in \ref{corpus}.
} (page title in Wikipedia case), 
the keywords of similar pages could be considered as synonyms for the keywords of the source page.
So the synonym search problem can be presented as a problem of similar pages search.
In turn, the issue of a similarity (between pages with hyperlinks) can be formulated as 
a problem of \emph{a search of similar graph vertices} based on hub and authority notion. 

\subsection{Problem statement}
\label{alg:problem:statement}

Given directed graph 
\(G=(V, E)\), the vertex set \(V\) (documents), the arc set \(E\) (links), the source vertex \(s\). 
For each document \(v\) there are two lists: \( \Gamma^+(v)\) 
includes documents which are referred by the source document \(s\), the list \( \Gamma^-(v)\) 
includes pages referring to the source document. For each vertex there are two weights, authority and hub:
\(\lbrace v \in V : a_v, h_v \in \mathbb{R}\rbrace\). 

It is needed to find the set \(A\) of vertices that are 
(i) \emph{authority vertices} to the source document \(s\) 
(i.e. value \eqref{eq:2a} for A is higher than for other subsets of vertices of the same size \(N\)), 
(ii) \emph{similar vertices} to the source \(s\) (i.e. there exists a vertices set \(H\) such that for each vertex from \(A\) 
there are vertices in \(H\) that simultaneously have out-links to \(s\) and to vertices from \(A\) \eqref{eq:2b}), 
(iii) the set \(H\) consists of hub vertices 
(i.e. value \eqref{eq:2c} for H is higher than for other subsets of vertices of the same size \(M\)). 
The aim is to select the set \(A\) of authority vertices and the set \(H\) of hub pages corresponded to \eqref{eq:2d}.
\begin{equation}
    \label{eq:2a}
   \sum\limits_{v \in A} {a_v} \xrightarrow[A \subset V, \lvert A \rvert = N]{} \max
\end{equation}
\begin{equation}
    \label{eq:2b}
    A \subset V, H \subset V, \forall a \in A ~~~ \exists h \in H : \Gamma^+(h) \owns \{s, a\}
\end{equation}
\begin{equation}
    \label{eq:2c}
    \sum\limits_{v \in H} {h_v} \xrightarrow[H \subset V, \lvert H \rvert = M]{} \max
\end{equation}
\begin{equation}
    \label{eq:2d}
    k \cdot \sum\limits_{v \in A} {a_v} + (1-k) \cdot \sum\limits_{v \in H} {h_v} \xrightarrow [A \subset V, H \subset V]{} \max, k \in [0,1]
\end{equation}

\subsection{Algorithm}
\label{alg:params}

The algorithm input parameters are 
\(s \in V\);
\({t, d, N, C_{max}}\in\mathbb{N}\);
\(\varepsilon \in \mathbb{R}\), where
\begin {itemize}
\item \(s\) -- the source document for which similar documents are sought for.
\item \(t\) -- root set volume (number of documents included in the root set of documents);
\item \(d\) -- the parameter defining base set\footnote{Base 
                   and root set forming is described below.}
               volume (\(d\) in-links for each document from the root set will be added to the base set, see more details in \cite{kleinberg99});
\item \(N\) -- number of similar words to be found (the same: number of similar documents to be found);
\item \(C_{max}\) -- maximum weight of the cluster 
(number of documents in the cluster, number of categories, etc. are taken into account), 
where cluster is a set of documents which have common hyperlinks to other documents and \emph{to categories};
\item \(\varepsilon\) -- the iteration error.
\end {itemize}

Steps of the adapted algorithm:
\begin {itemize}
\item [1.] The source document \(s\) refers to a set of documents 
that (along with \(s\)) form the root set of documents (\(\leq{t}\) documents should be included).
\item [2.] For each document in the root set of documents: all out-link documents and not more than \(d\) in-link documents are included in the base set (it is the same step as in HITS algorithm).
\item [3.] Document weights are calculated iteratively in accordance with HITS formulas. 
The iterations stop when the iteration error\footnote{Sum 
    of changes of hub and authority weights of all vertices after one iteration.} 
is less or equal to \(\varepsilon\).

HITS uses two values for rating pages: the authority value \(a_j\) and the hub value \(h_j\), 
which are defined in terms of one another in a mutual recursion 
(\(E\) is the set of hyperlinks):
\begin{equation}
    a_j = \sum\limits_{i:{(i,j)} \in E} h_i
    \label{eq:1a} 
\end{equation}
\begin{equation}
    \label{eq:1b}
    h_j = \sum\limits_{i:{(j,i)} \in E} a_i
\end{equation}
An authority value~\eqref{eq:1a} is computed as the sum of the hub values that point to that page. 
A hub value~\eqref{eq:1b} is the sum of the authority values of the pages it points to. 

\item [4.] Hierarchical clustering algorithm is applied to base set in order to cluster documents 
into group corresponding to different topics 
(this step is absent in Kleinberg algorithm). 
The clustering algorithm takes into account document weights 
(calculated on previous step) and hyperlinks structure of documents and categories.
\item [5.] For each cluster of the base set: select set \(A\) (contains \(N\) documents) such that 
(i) set \(A\) consists of authority documents 
(i.e. the total document authority weight is large enough compared to other subsets of documents of the same volume) and 
(ii) for each document \(a\) in \(A\) there are hub documents which refer to \(a\) and to the source document \(s\)\footnote{See 
    formulas \eqref{eq:2a}--\eqref{eq:2d} in problem statement, sec. \ref{alg:problem:statement}.
}.
\end {itemize}

\begin{figure}
    \centering
    \includegraphics[width=\linewidth]{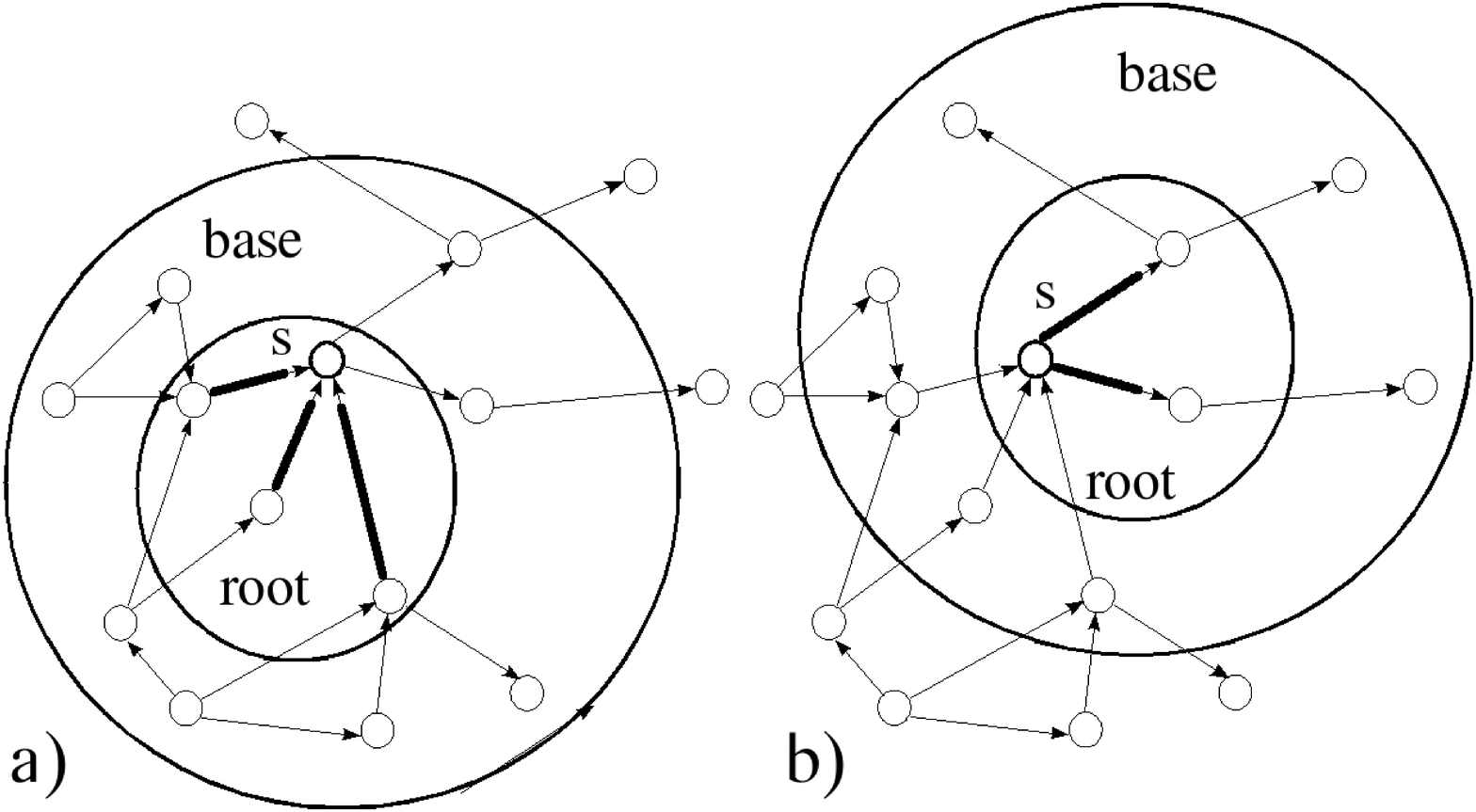}
    \caption{Root and base set forming in (a) HITS and (b) adapted HITS algorithm}
    \label{fig:baseforming}
\end{figure}

HITS and the adapted algorithms are different in the way of root set forming (\figref{fig:baseforming}). 
In HITS a root set consisting of \(t\) pages that \emph{point to} \(s\) (a) 
is assembled 
using relevant pages found by a search server (Altavista). 
In the adapted algorithm, pages, which are ponted by \(s\), form the root set (b);
then the root set grows into a base set as in HITS; 
and the result is a subgraph \(G\) in which hubs and authorities 
are identified\footnote{The 
    advantage of the approach presented in (\figref{fig:baseforming}a) is that 
    the root set includes hubs, so the base set can include authorities. The disadvantage is that 
    a lot of pages can refer to the source page \(s\), 
    root set volume should be constrained by \(t\), 
    and some hubs can be missed.

    In (\figref{fig:baseforming}b) number of out-links (the volume of root set) is limited, 
    since is constrained by size of the document \(s\). 
   But it is not evident that the base set will include authorities. 
   Experiments for comparison of the approaches are to be done (measures to compare results should be developed).
}.

The adapted algorithm also introduces the ideas of clustering base set of pages 
(into documents groups corresponding to different topics) 
 (i) using calculated hub and authority weights and 
 (ii) taking into consideration topics of the documents\footnote{See 
         the 3\(^{rd}\) characteristic of corpus documents in \ref{corpus}.
}.

\section{Implementation}

Adapted HITS algorithm was implemented in the program titled Synarcher.

Data for the program Synarcher are Wikipedia in MySQL format 
(the structure of tables corresponds to the requirements of the 
MediaWiki\footnote{\htmladdnormallink{MediaWiki} 
    {http://www.mediawiki.org} is a free software package
    originally written for Wikipedia.
}
system). 
Though Synarcher was tested with English and Russian Wikipedia data only, 
it is supposed that Synarcher can perform search within other wiki resources, which are based on MediaWiki, 
as well\footnote{See 
    \htmladdnormallink{sites using MediaWiki} {http://meta.wikimedia.org/wiki/Sites_using_MediaWiki}.
}.

The Synarcher was tested on Windows XP and Mandrake Linux. 
Running the Synarcher on client side requires a Java Runtime Environment (ver. 1.3.0 or above). 
The server should provide an access to Wikipedia resources via MySQL, Apache, and MediaWiki software. 
The currently public available Wikipedia servers (e.g. http://en.wikipedia.org) could not be used directly 
since the current version of Synarcher requires intensive computations. 
Hence a locally installed version of Wikipedia was used.

\begin{figure*}[ht]
    \centering
    \setlength{\fboxsep}{0cm}
    \fbox{\includegraphics[width=\linewidth]{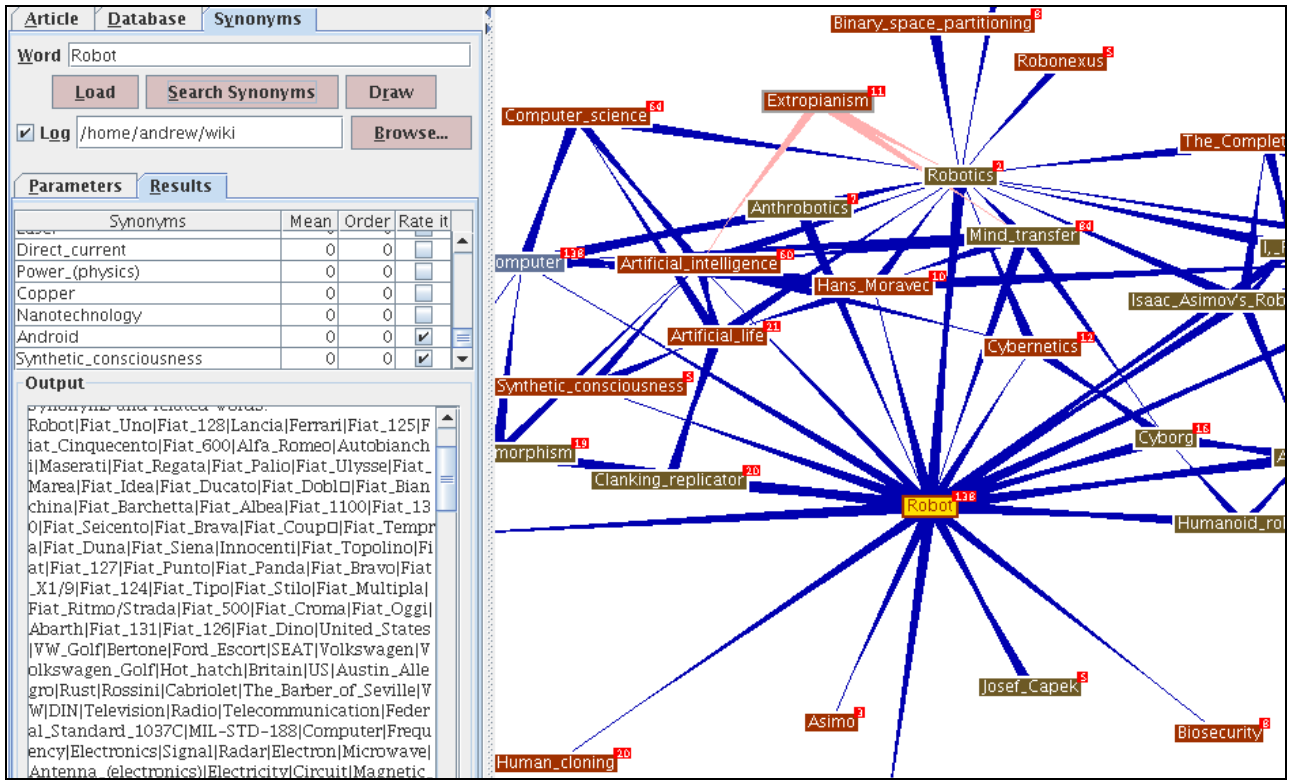}}
    \caption{The found synonyms and related terms (for the word Robot) presented in the table and graph (English Wikipedia)}
    \label{fig:robot}
\end{figure*}

The results of synonyms search are presented in the form of a table and graph (\figref{fig:robot}).
The graph presentation of synonym search results is based on TouchGraph WikiBrowser V1.02. 
The user is presented with a split view, with a conventional html browser 
on the left side of the screen, and a graph of a local region of the wiki on the right.

The link graph consists of nodes whose labels correspond to hypertext pages 
(labels are the names of articles in Wikipedia), 
connected by directed edges corresponding to hyperlinks between the pages. 
User can expand node, hide node's neighbours, or rate node as a synonym for the requested word.

The user enters a word, and the program performs automatic search of its synonyms. 
Then, the user explores the result graph (base set of articles for the source article \(s\)) 
and rates (option rate/unrate) words, which are synonyms from the user's point of view 
(this is an interactive part of the work with graph or table). 
Search parameters\footnote{See 
    parameters of algorithm in sec. \ref{alg:params}.} 
and synonyms selected by the user are stored at client side.

\section{Experiments}

The local versions of English and Russian Wikipedia 
(corresponding to the online version on March 8, 2005) were used in experiments. 
The English encyclopedia contains 901,861 pages, and 18,380,035 links, 
Russian encyclopedia contains 30,161 pages, and 468,771 links.

It should be noted, that the search for synonyms is not fully automated, 
since the program forms the list of related words, but many of words are not synonyms to the source word. 
The additional interactive search (in a graph or in a table) is needed.
Thus, the program provides filtering of related words (and potential synonyms), which serve as raw data for an expert.

The expert has identified (during an interactive search in Synarcher) seven synonyms for the word {\it robot}: 
{\it android, {\bf golem}, homunculus, domotics, replicant, sentience, parahumans}\footnote{Explanation 
    of the meaning of words can be found at \htmladdnormallink{} {http://en.wikipedia.org}.
}
(synonyms  presented in the thesauri 
\htmladdnormallinkfoot{WordNet} {http://wordnet.princeton.edu}
or 
Moby\footnote{\htmladdnormallink{Moby 
    Thesaurus List by Grady Ward} {http://www.dcs.shef.ac.uk/research/ilash/Moby}.
    }
are written here and below in boldface).
WordNet 2.0 contains only two synonyms for the word {\it robot}. 
They are {\bf automaton} and {\bf golem}. 

With the help of Synarcher four synonyms for the word {\it astronaut} were identified:
{\bf cosmonaut}, {\it taikonaut, spationaut}, {\bf space tourist}.
WordNet proposes two synonyms: {\bf spaceman, consmonaut}. 
There is no entry {\it robot} in Moby, 
but there are 79 related words for the word {\it astronaut} in it,
six of which can be considered as synonyms: 
{\bf aeronaut, cosmonaut, pilot, rocket man, rocketeer, spaceman}.
Thus, two and six synonyms for the words {\it astronaut} and {\it robot} were found, 
which are absent in thesauri WordNet and Moby.

Small size of the current version of the Russian Wikipedia does not allow one to find as many synonyms 
(within related terms gathered by Synarcher) as the English Wikipedia.

\section{Future work}

Future work includes estimation of the quality of synonym search, 
e.g., by comparing the found set of synonyms against the WordNet and Moby thesauri.

The hierarchical algorithm is planned to be used to cluster the result set of words 
related to different problem areas. 
It is possible for wiki resources, since every article has one or more assigned category types 
(e.g. the article "QNX" has the following categories in Wikipedia: Unix, Computing platforms, Embedded operating systems).

\section{Conclusions}

Adapted HITS algorithm for search of synonyms and related terms in the corpus with hyperlinks and categories is presented. 
The algorithm is implemented in the program Synarcher. 
English and Russian Wikipedia are used as the text corpus. 

The experiments show that successfull search of new synonyms is possible with Synarcher. 
New synonyms, which are absent in thesauri WordNet and Moby, for the words {\it robot} and {\it astronaut} were found. 
It could be explained by the properties of Wikipedia.\footnote{ 
    The English encyclopedia contains a lot of articles (998,470 in English on March 1, 2006) 
    related to different topics (science, art, politics, etc.). 
    It contains articles related to the modern topics, since the encyclopedia is updated every day.
}

The proposed algorithm could be applied to a query modification (expansion by synonyms) in a search engine. 
The program Synarcher can serve as an assistant in a synonym dictionary forming.

\section{Acknowledgements}
This work was partly supported by the Russian Science Support Foundation

\bibliographystyle{IEEEtran}
\bibliography{linguistics}
%
%
\end{document}